\documentclass{JINST}  
\usepackage{graphicx}
\usepackage{psfig}
\usepackage{epsfig}
\usepackage{subfigure}
\usepackage{amsmath,amssymb}
   \title{Cryogenic environment and performance for testing the Planck radiometers}
   \author{
L. Terenzi$^a$\thanks{Corresponding
author.}~, M. Lapolla$^d$, M. Laaninen$^b$, P. Battaglia$^d$, F. Cavaliere$^c$, A. De Rosa$^a$, N. Hughes$^e$, P. Jukkala$^e$, V. - H. Kilpi\"{a}$^e$, G. Morgante$^a$, M. Tomasi$^c$, J. Varis$^f$, M. Bersanelli$^c$, R.C. Butler$^a$, F. Ferrari$^d$, C. Franceschet$^d$, P. Leutenegger$^d$, N. Mandolesi$^a$, A. Mennella$^c$, R. Silvestri$^d$, L. Stringhetti$^a$, J. Tuovinen$^f$, L. Valenziano$^a$ and F. Villa$^a$\\
\llap{$^a$}Istituto di Astrofisica Spaziale e Fisica Cosmica, INAF\\ 
	via P. Gobetti, 101 -- I40129 Bologna, Italy\\ 
\llap{$^b$}Ylinen Electronics Oy\\
	Kauniainen, Finland\\
\llap{$^c$}Universit\'a degli Studi di Milano\\ 
	Via Celoria 16, 20133 Milano, Italy\\
\llap{$^d$}Thales Alenia Space Italia, Sede di Milano\\ 
	S.S. Padana Superiore, 290 -- I20090 Vimodtrone, Italy\\
\llap{$^e$}DA-Design Oy\\ 
	Jokioinen, Finland\\
\llap{$^f$}MilliLab, VTT Technical Research Centre of Finland\\ 
	Espoo, Finland\\
	Email:\email{terenzi@iasfbo.inaf.it}}

  \abstract
  {The Planck LFI Radiometer Chain Assemblies (RCAs) were calibrated in two dedicated cryogenic facilities.
In this paper the facilities and the related instrumentation are described.
The main satellite thermal interfaces for the single chains had to be reproduced and stability requirements had to be satisfied. Setup design, problems occurred and improving solutions implemented are discussed.
Performance of the cryogenic setup are reported.}


\begin{document}
%________________________________________________________________

\section{Introduction}

The Planck (\cite{tauber}) Low Frequency Instrument (LFI, \cite{mandolesi,LFI}) is an array of 22 pseudo-correlation radiometers, whose main sensitive devices are cryogenic HEMT-based low noise amplifiers located in the Front End Modules (FEMs). 
The signal coming from the sky is compared continously to the emission of a reference load (4K RL, \cite{valenzia}) in order to reduce the impact of gain fluctuations. The reference loads are small blackbodies connected to the outer shield of the High Frequency Instrument (HFI, \cite{lamarre}) and kept at a stable temperature of about 4K.\\
A flight-like low temperature environment is needed during ground calibrations \cite{villa_RCA,mennella_RAA} of the instrument.\\ 
An important requirement for the testing facilities is to provide the possibility to change, in a controlled way, the temperatures of some of the devices in order to ensure tests to be run in a correct way.\\
Radiometers' linearity and calibration constant \cite{mennella_LIS} are verified by changing the temperature of the reference loads and of dedicated sky simulators, while focal plane temperature variations effect on the radiometers \cite{terenzi_THF} is evaluated by changing the temperature of the front end modules. A stability control at 1\% level with respect to single temperature steps, typically ranging from 2 to 5 K, is then required for these main stages during calibration tests.\\
In this paper, the cryogenic facilities used for the ground tests of the LFI single Radiometer Chain Assemblies (RCAs) are described.\\
Two different chambers sharing the same design philosophy were used:\\ 
(1) the 30 and 44 GHz channel radiometers (\cite{lfi30}) were tested in the Thales Alenia Space - Italy (Milano) laboratories, while \\
(2) the 70 GHz channel radiometer chains (\cite{lfi70}) were tested in the DA-Design (Ylinen Electronics) laboratories, in Finland.

In Section 2 the thermal setup of the radiometer chain assemblies is described and an overview of the chamber used in Milano is given, specifically the mechanical setup, the main thermal interfaces design and the coolers used to obtain the desired temperatures, together with the temperature monitoring and control setup and a description of the sensor types and locations. The performance reached during the calibration tests in terms of temperature stability, together with some optimization study, is provided as well.\\
The main differences in the 70 GHz channel chamber are detailed in Section 3.\\
Finally, in Section 4, conclusions are addressed.

\section{Cryogenic facility setup for 30--44 GHz radiometers}

\subsection{The radiometer thermal scheme}

\begin{figure}[!ht]
	\centering
 \subfigure[]{\epsfig{file=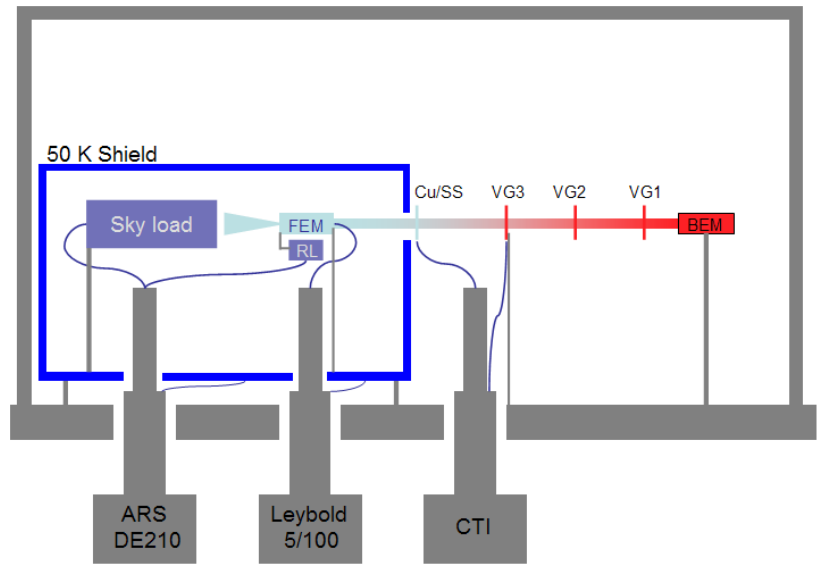,width=11.cm}} 
 \subfigure[]{\epsfig{file=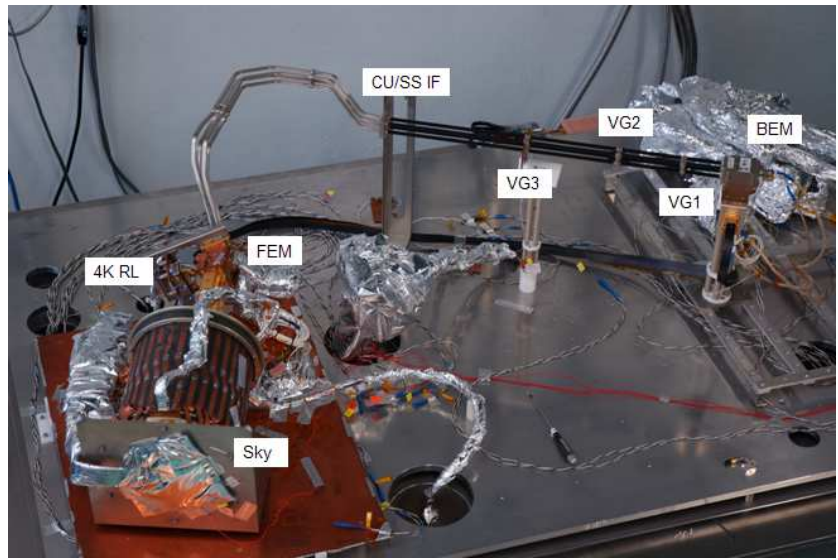,width=11.cm}}			
\caption{(a) A schematic view of the 30 and 44 GHz RCA thermal setup integrated in the cryogenic chamber. Main thermal interfaces are shown.
				 (b) Setup of a 30 GHz RCA, integrated in the chamber; insulating support of the BEM and third V-groove interface are visible.}
\label{rca_global}
\end{figure}

A single LFI radiometer chain spans a large range of temperatures from the Back End Module (BEM), at about 300 K room temperature, to the Front End Module at 20 K. They are connected by means of four long waveguides, about 1.7 meters long. The materials used for the waveguides allow a good inner signal transmission while having a good thermal insulation; in particular, the warmest part is built in stainless steel in order to create a large thermal gradient between the warm 300 K stage and the coldest stages.\\ 
At satellite level, three large radiative shields (the V-grooves) are interfaced to the waveguides and, in flight, will help them to dissipate heat loads to the surrounding 3 K environment.\\
A set of 4K reference loads is part of the LFI instrument, although mechanically and thermally linked to the HFI outer shield.
The loads, together with a dedicated sky simulator load, have to be taken into account in the design of a ground test cryogenic facility as well.
The overall thermal scheme of a Radiometer Chain Assembly, integrated in the facility, is shown in Fig. \ref{rca_global}a.

\subsection{The cryogenic chamber}

The size of the facility is about 2 $\times$ 1.2 $\times$ 1 ${\rm m}^3$. Such size is compatible with the volume of the radiometer chains, driven by the waveguides length and fine-shaping.\\

A copper shield at a temperature of about 50 K is used to radiatively insulate the coldest parts of the system, namely the FEM and horn at 20K,
the 4K reference load and a passive calibration tool (the {\it sky load}) simulating the sky microwave background emission, to be kept at about 4 K.\\
The chamber is provided with three cryocoolers and a vacuum bench allowing to reach pressures lower than $10^{-5}$ mBar.

\subsection{Thermal setup and main interfaces}\label{stages}

As mentioned above, the Planck LFI covers a large range of temperature stages, starting from about 300 K at the back end level down to the 
4 K reference load.\\
A set of thermal interfaces in the cryogenic chamber provides an optimized and stable working environment for the radiometers.\\
The seven main interfaces are listed here together with their working temperatures (see also Fig. \ref{rca_global}):

\begin{itemize}
 \item the back end module is controlled at a temperature of about 310 K;
 \item the third and coldest V-groove interface with the waveguides is controlled at a temperature of about 60 K;
 \item the copper-stainless steel interface of the waveguides is temperature controlled at 22 K;
 \item the shield surrounding the lower temperature zone is kept at about 50 K;
 \item the front end module is temperature controlled at 20 K;
 \item the 4 K reference load is controlled in a range 8-20 K;
 \item the sky load is temperaure controlled in the range 8-20 K.
\end{itemize}

\begin{figure}[ht!]
	\centerline{\epsfig{file=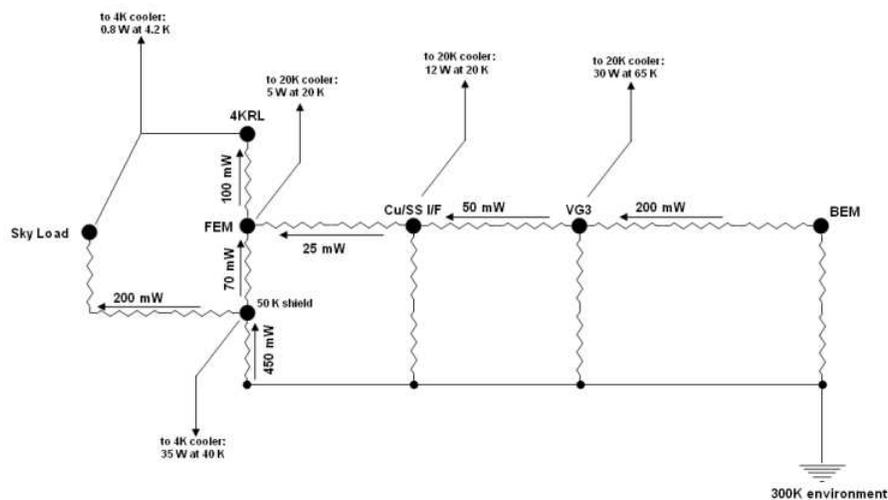,width=12.cm}}
	\caption{Thermal network of the chamber. Main thermal interfaces and heat loads are shown, together with cooling powers of the different coolers stages.}
\label{net}
\end{figure}

The possibilty to control and induce temperature changes is also needed in order to perform calibrations \cite{mennella_LIS} and check the temperature fluctuation effects on the instrument \cite{terenzi_THF,tomasi_DYN}.\\ 
Multilayer insulation (MLI) is extensively used in order to insulate from radiative heat load mainly the sky load large surface and the long waveguides path.
Due to the different routing of the RCA waveguides, different assembling and supporting structures were designed and built in order to get the optimal test setup. This led to some differences in the heat load and temperature control of some stages.
In Fig. \ref{net}, the thermal network and main heat routes, detailed in the following, are represented.

\subsubsection{The warm stage}

The mechanical support in the stages warmer than the cold shield is provided by two main structures (visible in Fig. \ref{rca_global}b), one supporting the BEM the other at the level of the third V-groove interface of the waveguides, and a further support provided by a kevlar wire holding the copper-steel interface of the waveguides. MLI is used to radiatively shield all the waveguides route.\\
The layout and materials used for the two main supports are the same for all the chains under test; four columns of PEEK polymer (half inch section for the BEM, quarter inch for the VG3 support) are alternated to small insulating bases of hostaform and teflon. The length of the supports is instead depending on the orientation of the waveguides and horns in the chamber.
This kind of insulation is enough for allowing a good control of the BEM temperature, since the latter is very close to room temperature.\\
The maximum passive load estimated on the VG3 interface, kept at 60 K, is about 0.2 W.\\
The thin kevlar wire linked to a stainless steel 304L support have a negligible impact (less than 1 mW) in terms of heat load on the interface, while warmer stages of waveguides contribute less than 50 mW.

\subsubsection{The cold stage}

The large shield covering the cold stage has a size of 0.72 $\times$ 0.53 $\times$ 0.60 ${\rm m}^3$.
It is made of copper and all its surfaces are covered with an MLI insulating layer.
Some mechanical details have been taken into account for the correct integration in the chamber.\\
The large dimension of the shield suggested to avoid too many mechanical boundaries in fixing the shield to the chamber base.
Only one of the supports, consisting in hostaform columns, is screwed to the base.
The others have an hollow base where a stainless steel ball sits.\\
This design reduces the stress due to thermal contraction during cooldowns while minimizing the contact surface to the chamber at room temperature. Another set of thin hostaform supports at its sides keeps the cold shield stable.\\
These solutions allow to keep the conductive load on the shield to less than 200 mW and the global load to about 450 mW, when taking into account also the radiative load on the large reflecting surface.\\
The 50 K shield contains, also radiatively insulating them from the room temperature walls, the coldest units of the LFI radiometers.\\
The front end module is connected by means of a stainless steel thin frame, alternated with teflon supporting bars, to the shield base. The supports are also holding the feed horn flange by means of a dedicated insert.\\
The purpose of these insulating supports is to keep the heat load on the FEM within 100 mW.\\
The reference load is mechanically supported by the FEM itself, by means of a similar dedicated stainless steel frame, in order to thermally insulate it from the 20 K stage. The conductive load is measured to be less than 100 mW.\\
The largest structure inside the cold part of the chamber is the sky simulator cylinder, whose diameter and length are about 25 cm and weight about 8 Kg. It is built in absorbing material (Eccosorb CR110) and its external surface is covered with MLI.
It is supported by a stainless steel AISI304 structure whose base is linked to the 40 K shield base by means of small hostaform bars. The global load on it is measured to be about 200 mW.\\

\subsection{The coolers}

The control of many temperature stages is required, in particular for the coldest temperatures which have to be varied during 
many of the calibration tests. Three two-stage coolers are then used to provide the right reference
temperature to the different parts of the instrument.

\begin{itemize}
\item	Both stages of a 20 K CTI Cryogenics cooler are used to keep the temperature of the 60 K V-groove and the 22 K stainless steel - copper waveguides interface at their nominal values.
\item	First and second stages of a further 20 K cooler (Leybold Coolpower 5/100) are used to keep the temperature of the radiative shield surrounding the cold part of the instrument at about 50 K and the temperature of the FEM at 20 K, respectively.
\item First and second stages of a 4 K cooler (Advanced Research System Inc. DE210) are used to keep the temperature of the 50 K radiative shield, as above, and the temperature of the reference and sky loads at about 8 and 10 K, respectively.
\end{itemize}

The coolers' model and cooling power are reported in Table \ref{cool_tab}. 
As expected, comparing their cooling powers to the heat loads on the various stages, the coolers have a good margin and then easily keep stages above 20 K at their nominal temperatures.

\begin{table}[h!]
\centering
	\begin{tabular}{l c c c}
	\hline
	Cooler Model 												& ARS  & Leybold  & CTI Cryogenics  \\
	 																& DE-210 & CoolPower 5/100 & Model 1020C \\
	\hline
	1st Stage Nominal T (K) 				& 41.0 			 & 80.0 									 & 65 \\
	1st Stage Cooling Power @ Nom. T (W) & 35 				 & 100 										 & 30 \\
	2nd Stage Nominal T (K) 				& 4.2 			 & 20.0										 & 20 \\
	2nd Stage Cooling Power @ Nom. T (W) & 0.8				 & 5											 & 12 \\
	\hline
\end{tabular}
\caption{Main characteristics of the coolers used in the cryogenic test chamber.}	
\label{cool_tab}
\end{table}

\begin{figure}[!h]
	\centering
 \subfigure[]{\epsfig{file=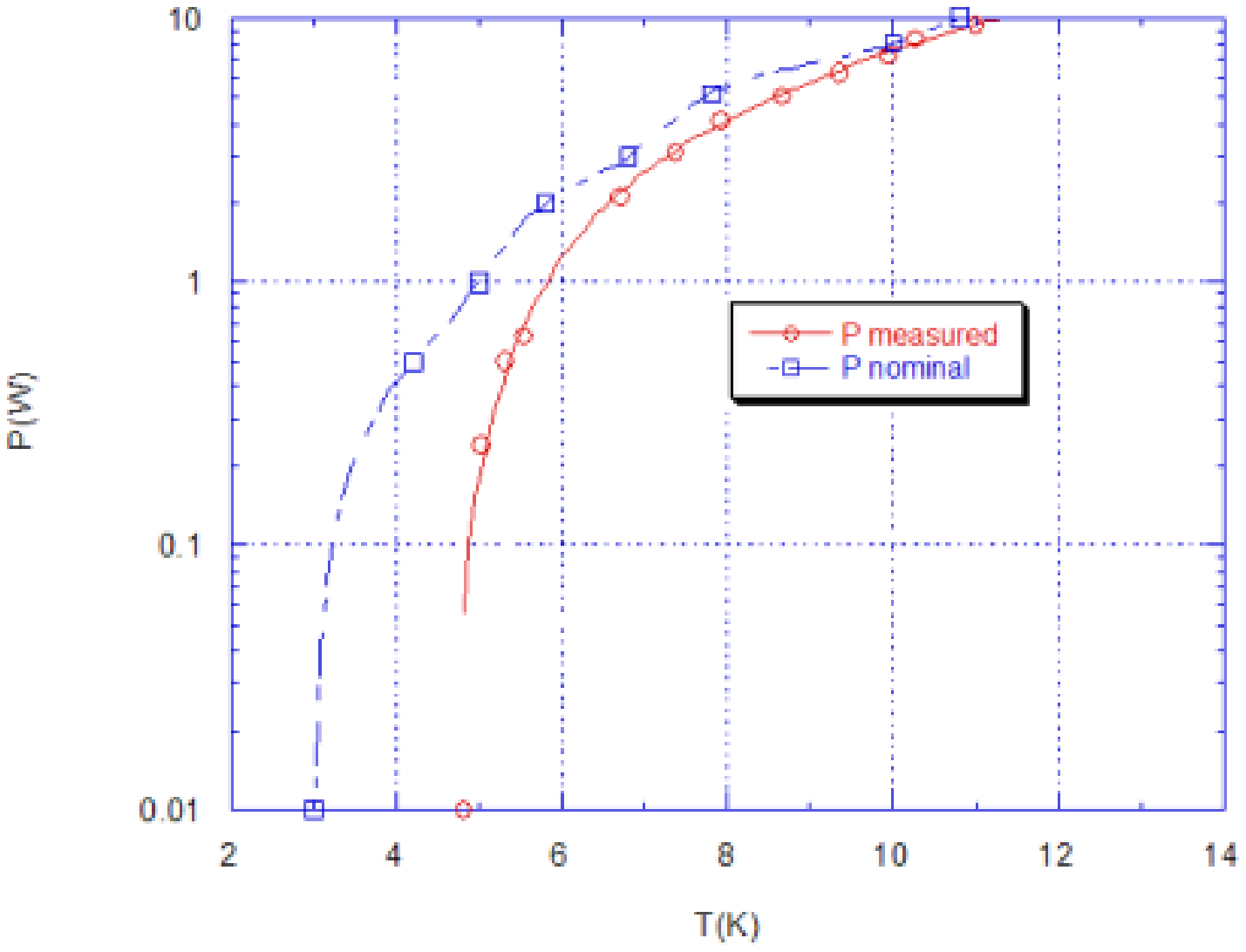,width=7.cm}} 
 \subfigure[]{\epsfig{file=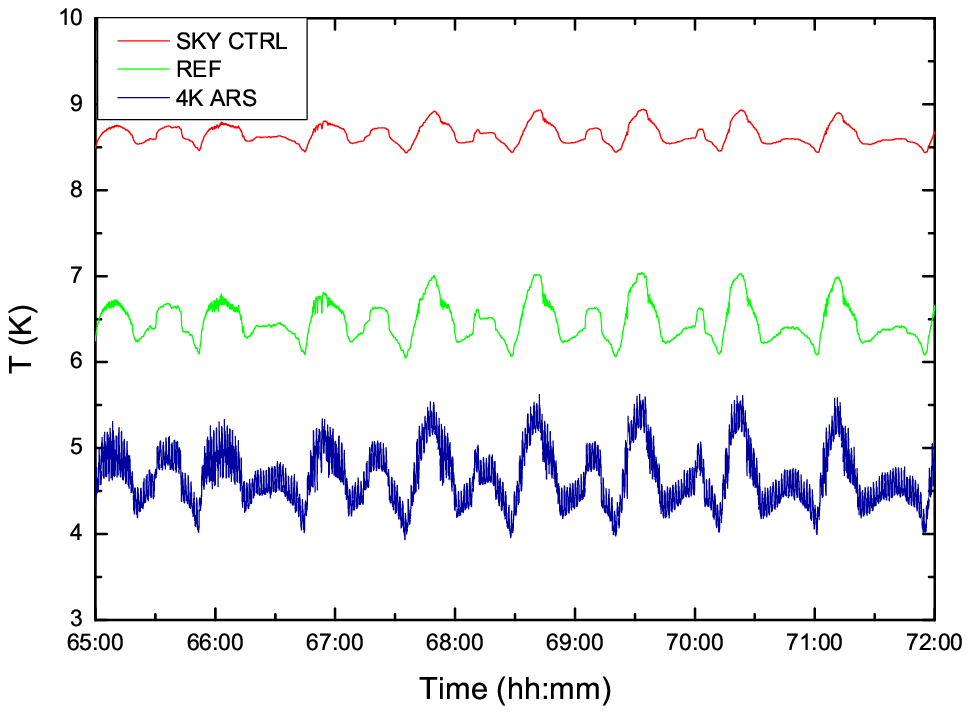,width=7.cm}}			
\caption{(a) Cooling power at different cold end temperatures: in the early setup (red curve), the cooler is far from the nominal (blue curve) behaviour, in particular at the lower temperatures. (b) This results in an unstable temperature sink for the reference and sky loads, which have to be controlled at temperatures higher than 8 K and 10 K, respectively.}
\label{before}
\end{figure}

Sky and reference loads are instead operated at temperatures higher than the nominal ones.\\
In the early runs, the 4 K cooler, actually, showed a high temperature instability (1 K peak to peak amplitude) and a cooling power lower than nominal (Fig. \ref{before}).

%\newpage

\begin{figure}[ht!]
	\centerline{\epsfig{file=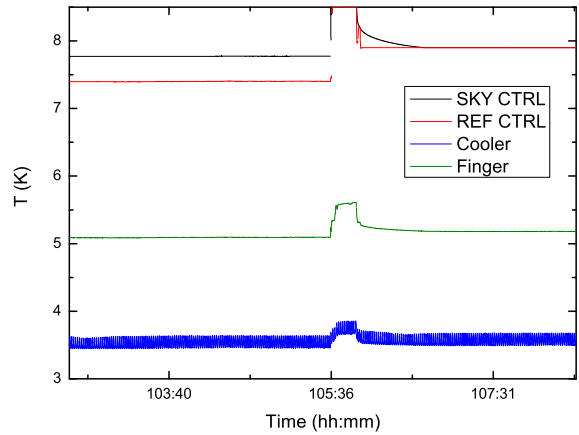,width=9.cm}}
	\caption{In the new setup the cooler cold head reaches nominal temperatures and the new cold finger inside the chamber is much more stable, 
	allowing a control of both loads at less than 8 K.}
\label{after}
\end{figure}

After some dedicated dry tests, a different configuration was decided for the cooler. It was mounted on the side of the chamber, with the cold finger pointing at the ground. An U-like thermal-vacuum path was built by means of a copper rod, which have become the new cold reference inside the cryogenic chamber. \\ 
As a result, the temperature during the tests is about 6 K, comparable to the former cold head one, but with a considerably improved stability.\\ 
As a consequence less power is needed for the stabilization of the loads (Fig. \ref{after}), keeping their temperature as low as possible.

This allows tests to be performed in almost nominal conditions.

\subsection{Temperature sensors, monitoring and control}

The large temperature range spanned by the instrument for the calibration tests needs an accurate monitoring and control.
A set of 18 temperature sensors is then used in the test campaign.\\ 
A couple of Lakeshore$^{\rm TM}$ PT-100 platinum resistance thermometers are used for the BEM control and monitoring close to the room temperature.
Silicon diodes (DT470 and DT670) and Cernox resistors with a sensitivity better than 1 mK are used for the low temperatures to control stages whose stability is relevant for a successful test campaign. \\
The temperature readout consists of two Lakeshore$^{{\rm TM}}$ LSC331 temperature controllers (two read out and control channels) and two LSC340 temperature controllers (one with two readout and control channels and 8 readout-only channels, the other with two readout and control channels and two readout only channels). These allows to control six of the seven stages listed in the Section \ref{stages}.
Every temperature stage, with the exception of the radiative shield, is actively controlled by heaters using PID algorithms.\\
Temperature controllers and monitors are interfaced, via GPIB boards, to a dedicated PC and driven by a dedicated software developed in the LabWindows environment, which also handles the communication with a breadboard model of the Data Acquisition Electronics (DAE) supplying the power to the radiometer devices and acquiring data from the detectors. Finally a quick look monitoring is performed by another software, RACHEL \cite{mala}, which also converts data in FITS format to prepare the off-line analysis \cite{tomasi_LIFE}.

\subsection{Thermal performance}

The main performance requirements for the cryogenic chamber are the cooling of the entire LFI radiometer chains down to the nominal low temperatures, allowing both the correct thermal stability and the correct temperature steps, when needed by the procedures.

\subsubsection{Temperature distribution}

During the whole flight model test campaign, the chamber setup and the temperature control has allowed a good repeatibility in the temperature distribution for the stages in the range 20-310 K, as reported in the Table \ref{hot_t}.

\begin{table}[h!]
\centering
	\begin{tabular}{c r}
	\hline
	Temperature stage & T (K) \\
	\hline
	FEM control &						   20.0 \\ 
	CU/SS Interface &				   22.0 \\ 
	BEM control &						   310.0 \\
	BEM body &	$\sim$    310.0 \\
	Third V-Groove & $\sim$ 60.0 \\ 
	Cold Shroud &	$\sim$    50.0 \\
	\hline
\end{tabular}
\caption{Typical set of temperatures obtained for the warmest stages of the RCAs.}	
\label{hot_t}
\end{table}

Due to the different orientation and routing of the feed horns and waveguides, different sky load support structures are used leading to different thermal paths between the 50 K shield and the load base. This leads to a heat load variation on the 4 K cooler, depending on the RCA under test.

In Table \ref{min_t}, as representative for the minimum controlled temperature reached for the sky and reference load backplates, values from noise properties data analysis are reported for each RCA.

\begin{table}[h!]
\centering
	\begin{tabular}{c c c}
	\hline
	RCA ID \# & Sky Load T (K) & Ref Load T (K) \\
	\hline
	24	&	8.5	&	8.5 \\
	25	& 10.5 & 8.0 \\
	26	& 13.3 & 8.0 \\
	27	& 12.8 & 9.5 \\
	28	& 8.6 & 8.5 \\
	\hline
\end{tabular}
\caption{Minimum stable temperatures reached for the noise properties test by the sky and reference load control backplates}	
\label{min_t}
\end{table}

\subsubsection{Temperature	stability}

Temperature instabilities are the most important thermal systematic effects for the LFI instrument also during calibrations.\\
The large number of temperature stages controlled during the tests is required to keep under control this dangerous source of systematics.
The design of the test and dedicated corrections adopted throughout the campaign ensure the high level of stability needed for many of the tests.\\ 
Typically, the short term peak-to-peak temperature variation is kept under a level of 5 mK for the lowest temperature stages.

\section{Facility setup for the 70 GHz radiometers}

\begin{figure}[h!]
	\centerline{\epsfig{file=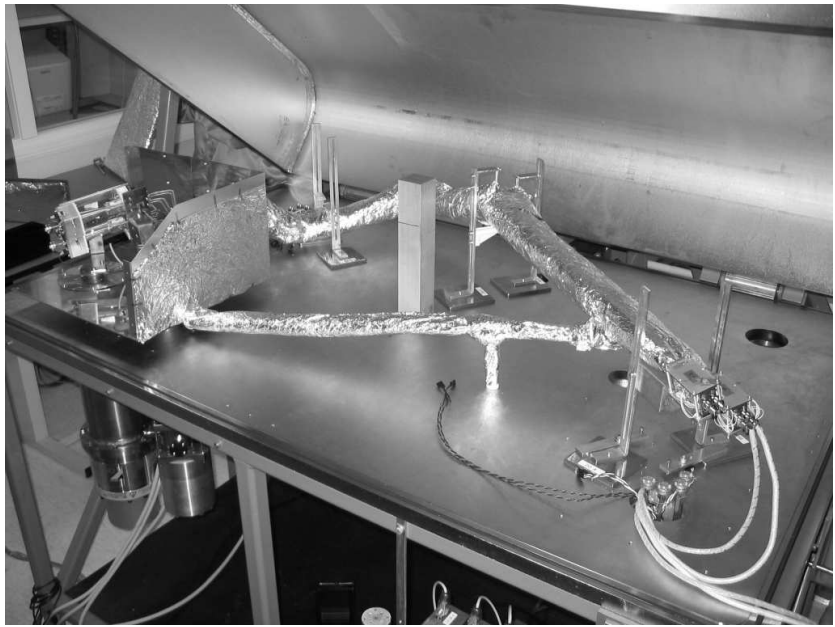,width=11.cm}}
	\caption{A global view of two RCAs integrated in the 70 GHz channel cryogenic chamber.}
\label{Finn_Chamber}
\end{figure}

The cryogenic facility has a volume allocation of about 1.6 $\times$ 1 $\times$ 0.3 $m^3$.
The design of the facility is very similar to the 30-44 GHz one, even though the 70 GHz facility is designed in order to house two radiometer chains at once (Fig. \ref{Finn_Chamber}). This is possible in terms of thermal budget, due to the smaller size of the radiometer units.\\ 
The heat transfer through the waveguides is reduced mainly by the effective section being about 0.45 times the section of the 30 GHz guides.
In terms of global envelope inside the cold shield, the sizes of the feedhorns and front end modules are reduced and the choice to use two small dedicated sky loads directly in front of the horns allows to use the same volume than for the 30 and 44 GHz one.

\subsection{Main thermal interfaces}

The chamber is implemented with less control stages than in the 30/44 GHz one and only two cryocoolers, used mainly for the coldest 4 -- 20 K stage.
Main interfaces are provided to both RCAs under test at once, by means of copper braids or slabs connecting them each other:

\begin{itemize}
 \item the thin rods of stainless steel supporting the back end modules insulate them from the chamber envelope; no control is implemented;
 \item the second V-groove interfaces with the waveguides are connected to the same interface of the cold stage shield at a temperature of about 60 K;
 \item the shield surrounding the lower temperature zone is kept at about 60 K;
 \item the front end modules are temperature controlled at 20 K;
 \item the 4 K reference loads are controlled in a range 10-20 K;
 \item the sky loads are temperature controlled in the range 10-25 K.
\end{itemize}

\begin{figure}[h!]
\centerline{\epsfig{file=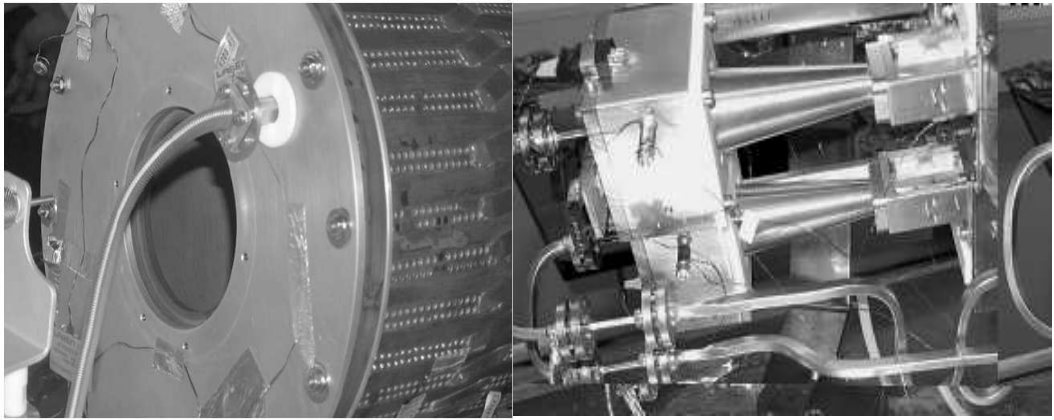,width=14.cm}}
\caption{The band shape is recovered sweeping the frequency of an input signal in the sky load. In the 30/44 GHz case the signal is injected 
through a flexible waveguide entering the sky load close to the feed horn aperture (left). In the 70 GHz the signal is injected from the back
of the sky loads (right)}
\label{swept}
\end{figure}

The power injection system on the sky loads, used to measure the radiometer bands \cite{zonca_SPR}, has a different setup in the two chambers.
In the 70 GHz, it is located on the back of the sky load in order to illuminate directly the feed horn, in front of it. In the 30 and 44 GHz tests, the power is injected by means of a wave guide mounted in the same face as the horn (see Fig. \ref{swept}) and the test was conducted by measuring the signal reflected by the sky load base to the feed horn aperture.

\subsection{The coolers}

Two two-stage coolers are used to provide reference
temperatures to the different parts of the instrument.

\begin{itemize}
\item	First stages of the 20 K cooler are used to keep the temperature of the second V-groove and of the radiative shield surrounding the cold part of the instrument at about 70 K.
\item	The second stage of the 20 K cooler is used to keep the FEMs temperature at 20 K.
\item The cold stage of the 4 K cooler are used to keep the temperature of the reference and sky loads at about 10 K.
\end{itemize}

\subsection{Temperature sensors, monitoring and control}

Temperature sensors are of the same type of those used for the Milano facility. In order to test two chains at the same time, it is preferred to dedicate two thermometers for some relevant stages, one for each RCA under test.\\
Two sensors are used to monitor the temperatures of each of the two sky load backplates and absorbers, and BEMs. Due to their smaller size, only one sensor is used to monitor the single interface common to the two FEMs and the single interface common to the two reference loads.\\ 
The acquisition is performed using two twin DAE breadboards of the same model used for 30 and 44 GHz test campaign.

\subsection{Thermal performance}

The main temperature stages stability is at the level of 10 mK within the typical duration of the test (about 1 hour).
The BEMs reach a temperature of about 305 K, while the FEMs are at their nominal 20 K temperature, with a stability better than 1 mK.\\
In the Table \ref{min_t}, the minimum controlled temperature reached for the sky and reference load backplates are reported for each RCA.

\begin{table}[h!]
\centering
	\begin{tabular}{c c c}
	\hline
	RCA ID \# & Sky Load T (K) & Ref Load T (K) \\
	\hline
	18 -- 23	&	11.4	&	9.7 \\
	19 -- 20	& 9.4 & 8.0 \\
	21 -- 22	& 9.6 & 7.7 \\
	\hline
\end{tabular}
\caption{Minimum stable temperatures reached for the basic properties test by the sky and reference load control backplates}	
\label{t70}
\end{table}

\begin{figure}[h!]
	\centering
 \subfigure[]{\psfig{file=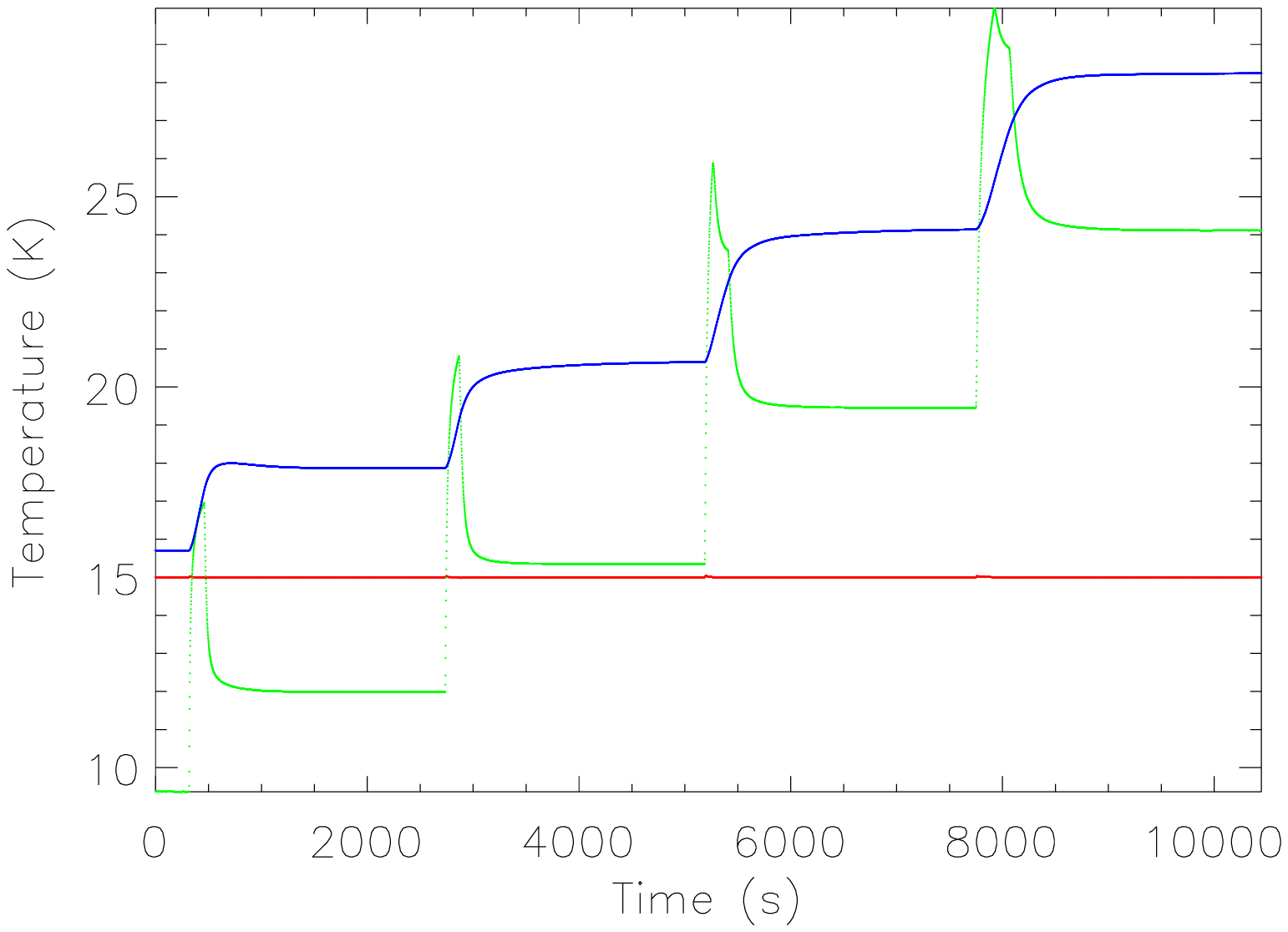,width=7.cm}} 
 \subfigure[]{\psfig{file=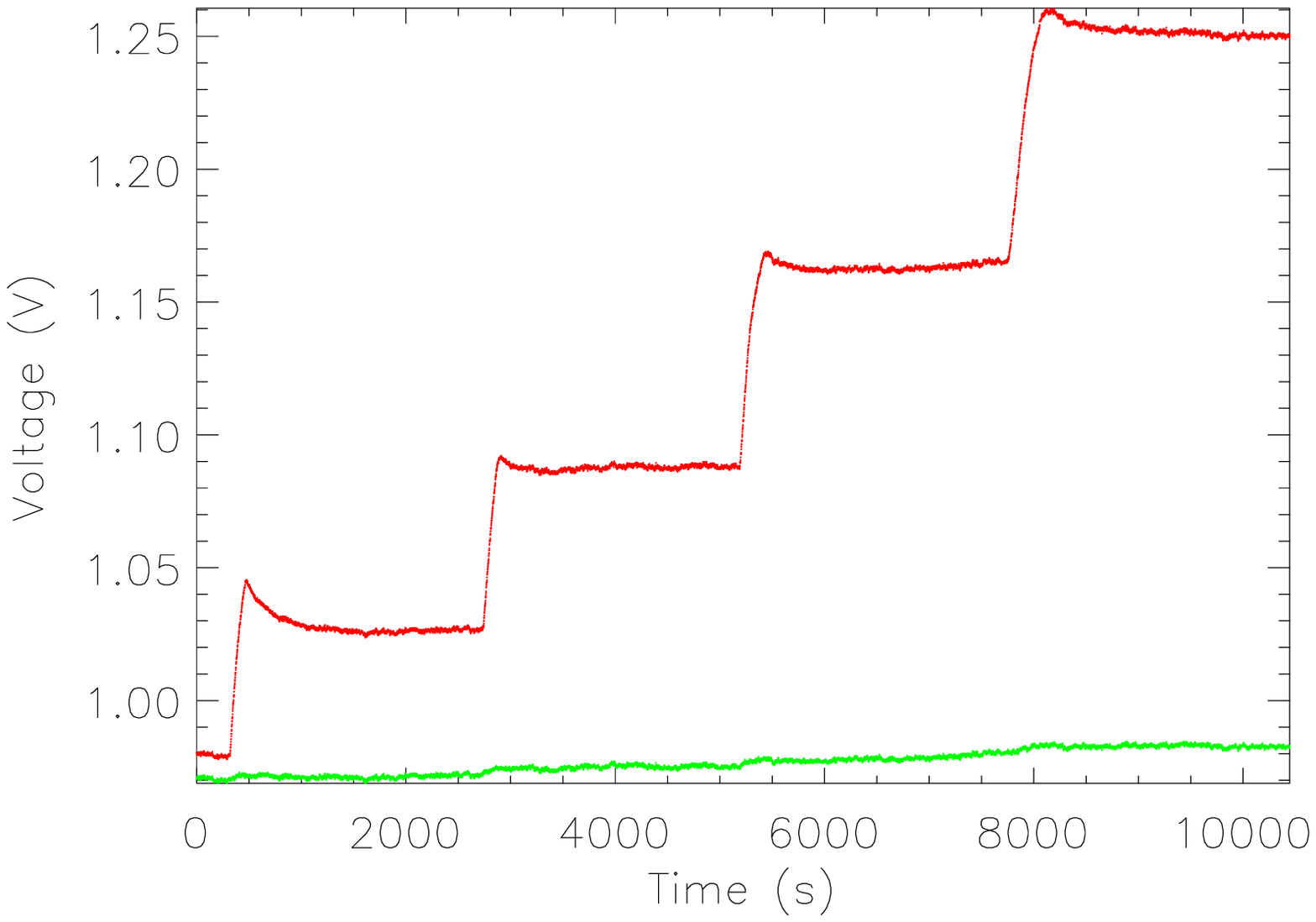,width=7.cm}}
\caption{RCA 19 channel A during sky load temperature steps test: (a) temperature curves for reference load (red), sky load backplate (green) and absorber (blue), (b) corresponding voltage output from one of the RCA 19 channels: sky output is in red reference output is in green. No sharp correlation is found between the temperature curves of the skyload sensors and the sky voltage output which shows an intermediate behaviour between the backplate and absorber temperature curve.}
\label{sky70}
\end{figure}

Two sensors are dedicated to each sky load, one on the back plate, while the other is inserted in a dedicated hole in the absorber part.\\
Some problems have occurred in the correct measurement of the temperatures of the small sky loads. There are systematically 4 to 9 K differences between the two temperature sensors. It is difficult to explain this large difference, also by modeling and correlating temperature data with radiometer outputs (Fig. \ref{sky70}).  

\begin{figure}[h!]
	\centering
 \subfigure[]{\psfig{file=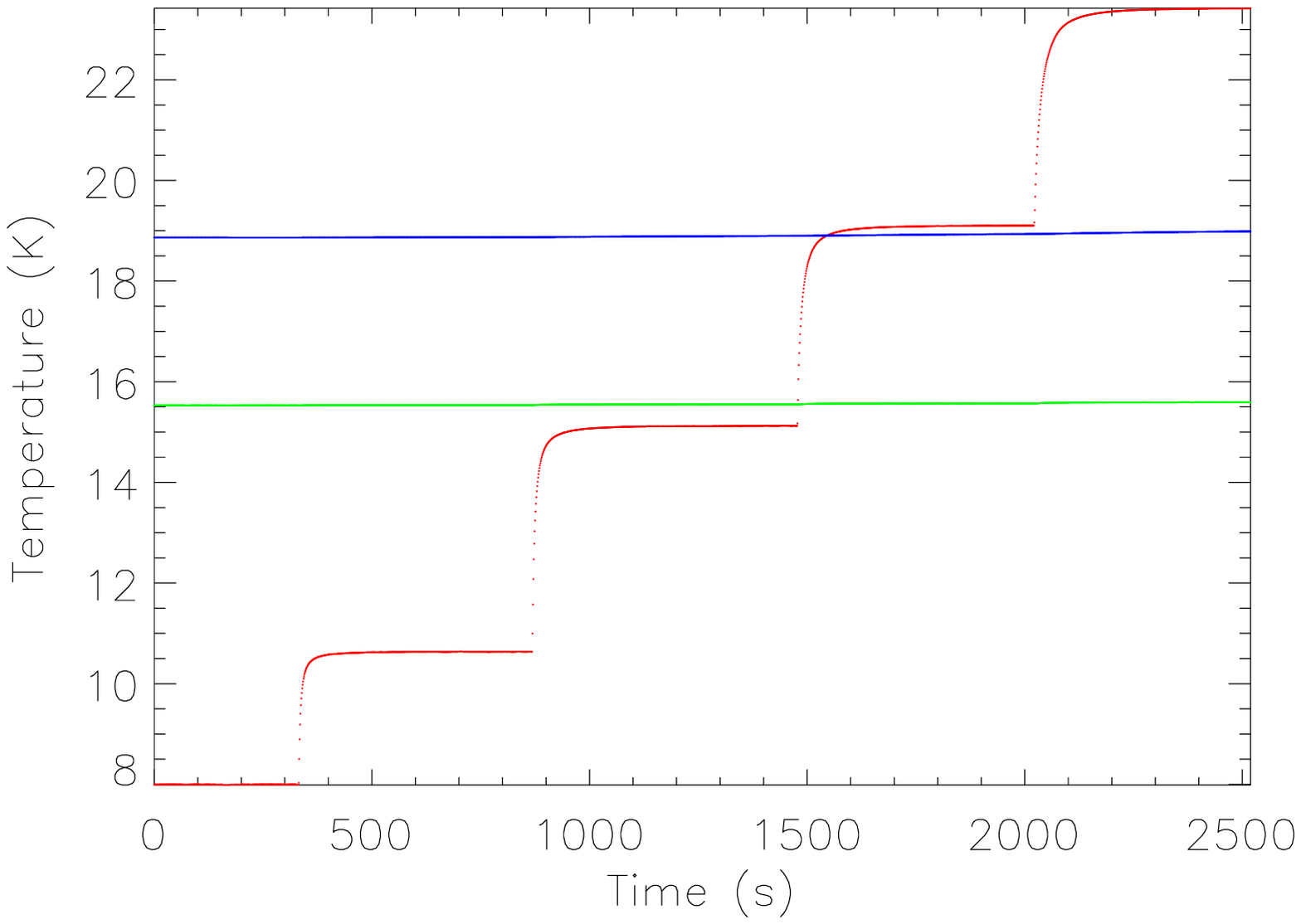,width=7.cm}} 
 \subfigure[]{\psfig{file=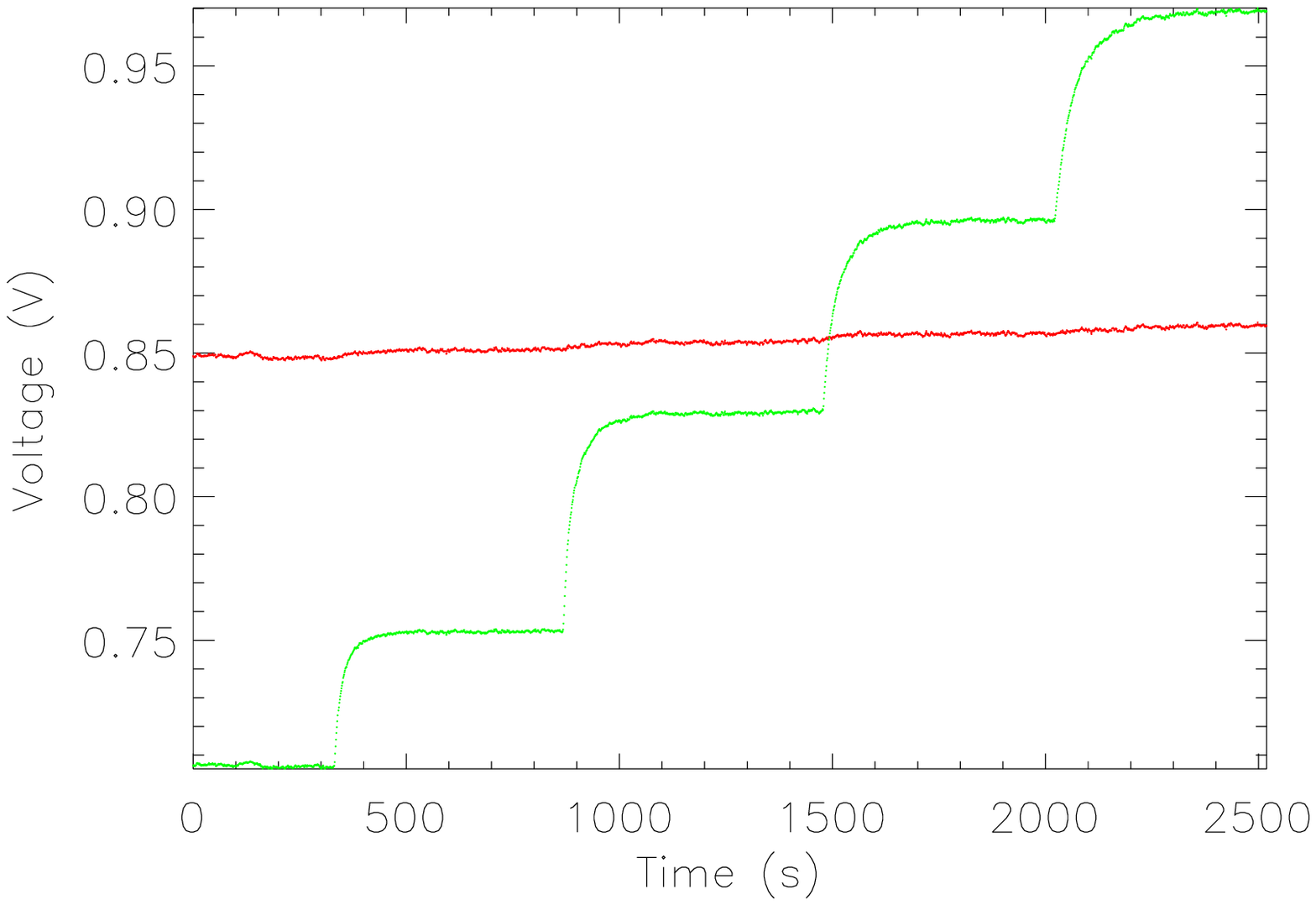,width=7.cm}}
\caption{Temperature curves of the skyload backplate and absorber and the voltage output from the RCA 20 channel A, during reference loads temperature steps: (a) temperature curves for reference load (red), sky load backplate (green) and absorber (blue), (b) corresponding voltage output from one of the RCA 20 channels: ref output is in green and shows a good correlation with temperature curve.}
\label{ref70}
\end{figure}

The main uncertainties in the temperature distribution within the sky loads derive from their mechanical assembly: the load is not bounded with the metallic envelop but it is simply blocked through the teflon tip. Due to differential contraction during cooldown, the thermal contact between the Eccosorb, which has a larger thermal contracion coefficient than aluminum, and its envelop could be reduced so to create a temperature gradient, as the measured one.\\
Due to these uncertainties, main results from the calibrations are derived from tests performed by changing reference load temperatures, whose correlation with radiometer data is more under control (Fig. \ref{ref70}).

\section{Conclusions}

The cryogenic environment equipments used for testing the Planck LFI Radiometer Chain Assemblies have been described.\\ 
Main requirements in terms of temperature stability of the main stages at the level of few mK are achieved.\\
Performance of the cryogenic facilities allows to run correctly the tests planned and obtain relevant results and important informations about the LFI flight model radiometers properties.

\acknowledgments

Planck is a project of the European Space Agency with instruments
funded by ESA member states, and with special contributions from Denmark
and NASA (USA). The Planck-LFI project is developed by an International
Consortium led by Italy and involving Canada, Finland, Germany, Norway,
Spain, Switzerland, UK, USA.
The Italian contribution to Planck is
supported by the Italian Space Agency (ASI).
In Finland, the Planck project was supported by the Finnish
Funding Agency for Technology and Innovation (Tekes).


\begin{thebibliography}{}

\bibitem{LFI} Bersanelli, M., Mandolesi, N., Butler, R.C. et al. 2009, A\&A, Submitted
\bibitem{lfi30} Davis, R.J., Wilkinson, A., Davies, R.D., et al., 2009, J-Inst, This issue
\bibitem{lfi70} Varis, J., Hughes, N., Laaninen, M. et al. 2009, J-Inst, This issue
\bibitem{lamarre} Lamarre, J.M, et al. 2009, A\&A, Submitted
\bibitem{mala} Malaspina, M., Franceschi, E., Bersanelli, M., et al. 2009, J-Inst, This issue
\bibitem{mandolesi} Mandolesi, N. et al. 2009, A\&A, Submitted
\bibitem{mennella_RAA} Mennella, A. , Bersanelli, M., Aja, B., et al., 2009, A\&A, Submitted
\bibitem{mennella_LIS} Mennella, A. , Villa, F., Terenzi, L., et al., 2009, J-Inst, This issue
\bibitem{morgante_RAA} Morgante, G., et al. 2009, J-Inst, This issue
\bibitem{tauber} Tauber et al., 2009, A\&A, Submitted
\bibitem{terenzi_THF} Terenzi, L., Salmon, M.,J., Colin, A., et al., 2009, J-Inst, This issue
\bibitem{tomasi_LIFE} Tomasi, M., Mennella, A., Galeotta, S., et al., 2009, J-Inst, This issue
\bibitem{tomasi_DYN} Tomasi, M., Cappellini, B., Gregorio, A., et al., 2009, J-Inst, This issue
\bibitem{valenzia} Valenziano, L., De Rosa, A., Cuttaia, F., et al., 2009, J-Inst, This issue
\bibitem{villa_RCA} Villa, F., et al., 2009, A\&A, Submitted
\bibitem{zonca_SPR} Zonca, A., Franceschet, F., Battaglia, P. et al., 2009, J-Inst, This issue

\end{thebibliography}
\end{document}